\magnification=1095

\hoffset+.25truein
\hsize=6truein \tolerance=10000
\voffset=.5truein
\vsize=8truein

\centerline{\bf FINITE TIMES TO EQUIPARTITION IN THE}
\vskip 3truept
\centerline{\bf THERMODYNAMIC LIMIT}
\vskip 18truept
\centerline{by}
\vskip 12truept
\centerline{\it J. De Luca$^{\rm (a)}$, A. J. Lichtenberg$^{\rm (b)}$,
S. Ruffo$^{\rm (c)}$}
\vskip 3truept
\centerline{(a) Instituto de F\'isica, Universidade Federal de S\~ao Carlos}
\centerline{Caixa Postal 676, 13565-905, S\~ao Carlos, SP, Brazil}
\vskip 3truept
\centerline{(b) Department of Electrical Engineering and Computer Sciences}
\centerline{and the Electronics Research Laboratory}
\centerline{University of California, Berkeley, CA 94720}
\vskip 3truept
\centerline{(c) Dipartimento di Energetica ``S. Stecco"}
\centerline{Universit\'a di Firenze, and INFN, 50139 Firenze, Italy}

\vskip 36truept

\baselineskip=24truept

\centerline{\bf ABSTRACT}

{\narrower\smallskip
We study the time scale $T$ to equipartition in a $1D$ lattice of
$N$ masses coupled by quartic nonlinear (hard) springs (the
Fermi-Pasta-Ulam $\beta$ model). We take the initial energy to be
either in a single mode $\gamma$ or in a package of low frequency
modes centered at $\gamma$ and of width $\delta \gamma$, with both
$\gamma$ and $\delta \gamma$ proportional to $N$. These initial
conditions both give, for finite energy densities $E/N$, a scaling
in the thermodynamic limit (large $N$), of a finite time to
equipartition which is inversely proportional to the central mode
frequency times a power of the energy density $(E/N)$. A theory of
the scaling with $(E/N)$ is presented and compared to the numerical
results in the range $0.03\leq E/N \leq 0.8$.

\smallskip}
\vfil\eject

\centerline{\bf I.\quad INTRODUCTION}

In previous work [1,\thinspace 2], the time scale to achieve
equipartition in a nonlinear lattice of masses coupled by hard springs
(FPU-$\beta$ model) was studied with the initial energy primarily 
in a low-frequency mode, of mode number $\gamma$. It was shown in [1],
numerically and theoretically, that energy transfer to high frequency
modes is exponentially slow in a perturbation (energy) parameter at
low energy. The mechanism of a transition to more rapid energy
transfer is that for energy above a threshold $E\propto N^{-1}$,
interaction of neighboring low frequency modes will lead to a local
superperiod beat oscillation, of period
$T_{B}\propto N^{2}/\gamma E_{\gamma}$, that is stochastic. At a
sufficiently high energy a resonance with high frequency modes leads
to a transition to a fast diffusion regime which occurs above a
critical energy $E=E_{c}$, independent of $N$. The equipartition
time scale for $E> E_c$ was studied numerically in [1,\thinspace 2],
with the more detailed numerical investigation of the scaling time
to equipartition in [2] for $0.2 \leq E/N \leq 1.0$ giving
$T\propto T_B \sqrt{N}$. The $\sqrt{N}$ was interpreted as a 
size-dependent correction to the $T_B$ time scale. In a heuristic 
treatment of a more general oscillator chain we indicated a possible 
explanation for the $\sqrt{N}$ mode filling factor [3]. If we 
additionally set $\gamma \propto N$ in $T_B$, above, then $T_B$ only 
depends on the total energy per mode (energy density). The numerical 
results in [2] on the other hand, indicated that $T$, having an 
additional $\sqrt{N}$ dependence, would become infinite with $N$ at 
constant energy density, and therefore $T$ would not have a finite
value in the thermodynamic limit.   

In another study [4], the energy was placed in a low frequency
package centered on a mode $\gamma \propto N $ and with an extension
$\delta \gamma \propto N$. In that work the numerical results
indicated that the equipartition time was dependent only on the energy
density $E/N$ and therefore remained finite in the thermodynamic
limit, for a finite $E/N$. However, the measure of equipartition used
in that study was rather insensitive, so that the exact scalings were
difficult to obtain.

Other work has also discussed the question of time to equipartition
for various initial conditions and nonlinearities [5--10]. In an
early investigation, at higher energies, the relation between time
scales and Lyapunov exponents was investigated [5].  At the higher
energies investigated in that study, there was primary mode overlap,
for which a weak power law behavior was observed, with a transition
to stochasticity governed by a random matrix approximation. 
The existence of weak chaos at vanishing energy was further explored
in Ref. [11]. In other studies of more regular systems, such as
the discretized integrable
sine Gordon equation [6], and the FPU-$\alpha$ cubic nonlinearity,
which is a two term Taylor series approximation to the integrable
Toda lattice [8], abrupt transitions were found between power law
behaviour and ``essentially integrable'' behaviour. This class of
coupled systems is well worth further study, but has significantly
different properties from the more thoroughly studied FPU-$\beta$
system. In another study a $\phi^4$ nonlinear chain was introduced to
allow the transition between power law and exponentially slow
behavior to be studied [9]. The attention was mainly on the
exponentially slow time scales in the low energy (small quartic
nonlinearity) region, which is not our concern here. In
this paper we examine the relationship between initial conditions
for which the energy is placed primarily in a single mode $\gamma$,
which we call ``mechanical'' initial conditions, with initial condition
for which the energy is placed in a finite $\delta \gamma$ package of
neighboring modes with $\delta \gamma \propto N$, which we call
``physical'' initial conditions. We find,
numerically, that there is a transient time $t\propto \sqrt{N}$, with
mechanical initial conditions, which does not exist or is of minor
importance for the physical initial conditions. This transient
behaviour manifests itself at relatively small $N$ (relatively large
$E/N$), explored in [2] and [3] but tends to disappear for large $N$,
thus preserving the thermodynamic limit.    

The FPU-$\beta$ model is a chain of $N$ masses, coupled to nearest
neighbors by hard quartic nonlinear springs. The Hamiltonian
representing the chain is
$$
H=\sum_{i=0}^{N}{{p_i^2} \over {2}}+{1 \over 2}
(q_{i+1}-q_{i})^{2}+{{\beta} \over {4}}(q_{i+1}-q_i )^4. \eqno(1)
$$
We consider the case of strong springs $(\beta >0)$ and fixed
boundaries $q_0=q_{N+1}=0$. The constant $\beta$ describing the
strength of the anharmonic potential can be scaled to any positive
value. We vary the energy and fix $\beta$ at the value 0.1 to compare
with previous studies of the FPU lattice. The equations of motion are
integrated using a fourth order symplectic integrator. The harmonic
part of the Hamiltonian can be put in the form of $N$ independent
normal modes via the canonical transformation 
$$
q_{i}=\sum_{j=1}^{N}u_{ij}Q_j\quad j=1,N, \eqno(2)
$$
with canonical variables $Q_j$. The columns of the matrix $u_{ij}$
are the orthonormal eigenvectors of the positive definite Hermitian
eigenvalue problem for the $Q$'s. The frequencies $\omega_j$ of the normal
modes $Q_{j}$ are 
$$
\omega_j=2\sin \left( {{\pi j} \over {2N+2}} \right). \eqno(3)
$$
The above transformation puts the Hamiltonian (1) into the form 
$$
H=\sum_{i=1}^{N} \left( {1 \over 2} P_i^2 + {{\omega_i^2} \over {2}}
Q_i^2 \right) + {{\beta} \over {(8N+8)}}\sum_{i,j,k.\ell}
\omega_i \omega_j \omega_k \omega_l G(i,\,j,\,k,\,\ell )
Q_i Q_j Q_k Q_{\ell}, \eqno(4)
$$
where the coefficients $G$, as calculated in [1], are 
$$
G(i,\,j,\,k,\,\ell )=\sum_{P}B(i+j+k+l), \eqno(5)
$$ 
where $P$ represents the eight permutations of sign of $j$, $k$ and
$l$ and the function $B(x)$ takes the value 1 if the argument is
zero, -1 if the argument is $\pm 2(N+1)$, and zero otherwise.

\centerline{\bf II.\quad NUMERICAL CALCULATIONS}
 
The main numerical tool we use is the calculation of the effective
number of modes $n_{eff}$ containing the energy. We use the general
formalism of our previous work [1,\thinspace 2] that the linear
energies $E_{i}\equiv {1 \over 2} \left( P_i^2 + \omega_i^2 Q_i^2 \right)$,
$i=1,\ldots ,N$, are calculated as a function of time. The information
entropy is given by $S=-\sum_{i=1}^{N}e_i\ln e_i$, where
$e_i=E_i/\sum_{i=1}^N E_i$ are the normalized energies. We define the
effective number of modes sharing the energy by
$n_{eff}(t)\equiv \exp S$ [1--3]. We divide $n_{eff}$ by $N$, to get
a fraction in the range from zero to one, which we plot versus time
for various values of $N$ and energy densities $E/N$. We also average
over 10-20 different realizations of the initial mode phases to
minimize fluctuations. We take care to distribute the phases of the
modes in a random way so that the quartic term in (1) does not make
the total energy very different from the linear energy. In this way 
one is always close to a set of slightly perturbed linear oscillators
as long as $(\beta E/N) {\buildrel < \over \sim} \,\, 1$.

For ``physical'' initial conditions we distribute the total energy
$E$ uniformly among $\delta \gamma=N/16$ low-frequency modes ranging 
from $N/64$ to $5N/64$ with randomly chosen phases, for a fixed value
of $E/N$ over a range of $N$, and plot $n_{eff}/N$ versus time. We
compare the results to `` mechanical'' initial conditions in which
energy is placed in a fixed number of modes where $\delta \gamma=5$
with the modes ranging from $N/64$ to $(N/64)+4$. We have worked with
other values of $\delta \gamma$, finding the same qualitative results.

In Figure 1 we show the evolution of $n_{eff}/N$ at fixed energy
density $E/N=0.1$ for $N$ ranging from 256 to 4096 for
$\delta \gamma=N/16$. The data are seen to lie essentially on a
universal evolution curve with the correspondence improving for large
values of $N$ (error bars, where shown, refer to the statistical
error computed over the different initial phases; otherwise errors
are of the size of the symbols). This result is verified numerically 
for energy densities
in the range of $E/N$ $[0.03,0.8]$. In Figure 2 we show the evolution
of $n_{eff}/N$ for the initial $E/N=0.05$ and for the same range
of $N$, as in Figure 1, but for $\delta \gamma=5$ and centered around 
$\gamma=(N/64)+2$. We see that there is an initial transient with the 
evolution of $n_{eff}/N$ versus time, different for each $N$-value, but 
coalescing later in the evolution at longer times and higher
values of $n_{eff}/N$. If we introduce a factor of $\sqrt{N}$ to
normalize the time scale (not shown), we then find that the evolution
curves coalesce at early times but then diverge. These results are
qualitatively consistent with the numerical study in [2] (Fig. 5),
which indicated an extra volume filling factor, proportional to
$\sqrt{N}$, when the energy was placed primarly in a single low
frequency mode, typically $\gamma=3$. In this case the primary driving 
frequency is a beat $\Omega_B\propto \gamma E/ N^2$, with $\gamma=3$, 
such that the time for transferring energy is much longer than in the 
present situation. This allows the filling of the low frequency modes
by successive excitations (see ref. [1]) to manifest itself in the
additional $\sqrt{N}$ dependence. In Figure 3 we plot $n_{eff}/N$ as a
function of the logarithm of time, for a range of $E/N$ values
again with ``physical" initial conditions (note that $n_{eff}/N$
asymptotically converges to a value which is smaller than 1, due
to fluctuations as computed in Ref. [6]).
The evolution is a monotonically increasing function with an initial
transient and later an approximately linear increase on the
logarithmic time scale. The linear part shifts to the left by
somewhat less than a decade with every doubling of the energy
density, which indicates a power law increase of the time scale with
$(N/E)^\alpha$ having an approximate exponent $\alpha \simeq 3$. To
be more quantitative we will normalize the time by a function of
$N/E$, as in [2] (Fig. 5), to determine if all values of $N/E$ can
be fitted on a universal curve, after estimating the value of
$\alpha$ analytically.
\vskip 6truept
\centerline{\bf III.\quad A SCALING ESTIMATE AND NUMERICAL COMPARISON}
 
In the following we present an approximate theory of Hamiltonian
diffusion to explain, qualitatively, the power law behavior at low
energy densities. We start by assuming that there is an effective
number of low-frequency modes $\delta k$ that are responsible for
stochastically transferring energy to the high-frequency modes, leading
to equipartition. We assume that the initial mode package containing the
energy, $\delta \gamma$, is such that $\delta \gamma \leq \delta k$,
where the size of the effective package $\delta k$ determines the
couplings to high frequency. Because, for early times, most of the
energy is in the low-frequency modes, it is convenient to classify
the quartic monomial appearing in the sum of (4) depending on how
many of the four Q's in it belong to low-frequency modes. The largest
quartic terms at early times have the four Q's of low-frequency
modes. These couplings produce deformations of the linear actions of
the low frequency modes, creating stochastic separatrices for those
modes. In [1] it was shown that the necessary energy to create
separatrices in the low frequency modes has $E\propto 1/N$ if
energy is placed in a single mode. If we place energy in
$\delta \gamma \propto N$ modes we expect stochastic separatrices
to be created for $E/\delta \gamma \propto 1/N$, such that $E$
is independent of $N$. However, this occurs at low energy for
$\delta \gamma /N$ small.

Because of the nonlinear couplings among the low-frequency modes, the
frequency of these modes is corrected by a beat that we approximate
in the following way. We substitute canonical action angle variables
$Q_i=\sqrt{(2I_i/\omega_i)}\cos(\phi_i)$ and
$P_i=\sqrt{(2\omega_i I_i)}\sin(\phi_i)$ into the Hamiltonian (4),     
to obtain
$$
H=\sum_i \omega_i I_i + \left( {{\beta}\over{2N+2}} \right)
\sum_{i,j,k,l} G(i,j,k,l)
\sqrt{ \omega_i \omega_j \omega_k \omega_l I_i I_j I_k I_l}
\, ang (ijkl), \eqno(6)
$$
where $ang (ijkl)\equiv \cos(\phi_i)\cos(\phi_j)\cos(\phi_k)
\cos(\phi_l)$. The frequency of
mode $i$ is the derivative of the Hamiltonian respect to $I_i$,
which evaluates to
$$
\Omega_i=\omega_i +\left( {{\beta}\over{N+1}} \right)\sum_{j,k,l}
G(i,j,k,l)\sqrt{ \omega_i \omega_j \omega_k \omega_l I_j I_k I_l}
\, ang(ijkl)/\sqrt{I_i}. \eqno(7)
$$
We further assume that there is a rapid spreading over low frequency
modes, as observed numerically [1], such that we are considering the
sum to run over some $\delta k$ modes, to be determined. After using
the selection rule (5), the number of terms in the above sum is then
of the order of $(\delta k)^2$. We also assume every quartic term
in this sum is typically of the same size, i.e. with equal energies
for all low frequency modes, $\omega_i I_i=E/\delta k$. Since these
terms come with random phases, according to standard
gaussian statistics we take the sum to be proportional to the square
root of the number of terms. With these assumptions, and setting
$\omega_i = \omega_j$ so that $(I_j / I_i)^{1/2}$ cancels, (7) becomes
$$
\Omega_i \approx \omega_i + \omega_i
{{\beta E} \over {N}} \eqno(8)
$$
with the $\delta k^{-1}$ in the energy per mode cancelling the
$\delta k$ effective couplings. In (8) and below we set $N+1 \simeq N$
(large N) except where it appears in the selection rule.
>From (3) $\omega_i \simeq \pi i/N$, for low frequency
modes, and taking the  beat frequency $\Omega_B$, between $\omega_i$
and a neighboring mode, to be of the order of the nonlinear frequency
shift, we obtain, for any $i$ (with $i \ll N$),
$$
\Omega_B \sim {{\pi i} \over {N}} \, {{\beta E} \over {N}}. \eqno(9)
$$

The change in the linear energy $E_i={1 \over 2}(P_i^2+\omega_i^2Q_i^2)$
of a driving low frequency mode $i$, can be calculated by taking the
derivative of (6) with respect to the angle $\phi_i$ 
$$
{{dE_i} \over {dt}} =- \left( {{2 \beta} \over {N+1}} \right) \omega_i
\sum_{j,h1,h2} G(i,j,h1,h2)
\sqrt{ \omega_i \omega_j \omega_{h1} \omega_{h2} I_i I_j I_{h1} I_{h2}}
\sin \phi_i \, ang (j,h1,h2), \eqno(10)
$$
As in our previous work [1] the notation $h1$ and $h2$ explicitly
indicates that the energy transfer occurs only between low frequency
beat oscillation and high frequency mode difference oscillations
through the Arnold diffusion mechanism.
In the above equation, the summation is over indices $j$, $h1$ and
$h2$ for a given $i$. The only terms to transfer energy to high
frequency modes are the ones where $j=i$, since then the product of
the two low frequency angles does not have a fast phase associated
with it. Additionally, the selection rule requires that $G(i,j,h1,h2)$
will be zero unless
$$
2i+h1+h2=2N+2, \eqno(11)
$$
for which $G= -1$. Expanding the dispersion (3) at high frequency and
using (11),
$$
\delta \omega_h \simeq (\pi^2 i/2N^2 )(h1-h2). \eqno(12)
$$
In order for the low frequency beat oscillations to transfer energy
by Arnold diffusion to the high frequency beats we require
$\Omega_B\,\, {\buildrel > \over \sim} \,\, \delta \omega_h$. From (9)
and (12) the inequality gives
$$
{\pi \over N} \,\, i \,\, {{\beta E} \over {N}}
\,\, {\buildrel > \over \sim} \,\,
{{\pi^2} \over {2N^2}} \,\, i \,\, (h1-h2)_{\rm Max}
$$
which reduces to
$$
\delta h \equiv (h1-h2)_{\rm Max} \,\, {\buildrel < \over \sim} \,\,
{{2 \beta E} \over {\pi}} \eqno(13)
$$
To determine the number of terms in the sum in (10) we note that for
every $i$ we can take $h2$ arbitrarily from the high frequency
package of $\delta h$ modes and then $h1$ is calculated from (11)
with the restriction, from (13) that
$h1-h2 \,\, {\buildrel < \over \sim} \,\, 2\beta E/\pi $.
Writing (11) in the form
$$
i = N+1-h_2 - \ell /2 \quad , \quad \ell \,\, {\rm integer},
$$
such that for $i=1$, we have
$$
\eqalign{
h_2 = N&-1,\qquad \ell = 2 \cr
h_2 = N&-2,\qquad \ell = 4 \cr
&\vdots\cr
}
$$
up to $i=\delta h$ for which
$$
h_2 = N - \delta h/2, \qquad \ell = \delta h .
$$
Thus we have a decreasing number of couplings with increasing $i$,
with the average number of couplings per low frequency mode
$\delta h/4$ which scales with $\beta E$ as in (13). Substituting
this result, together with (13), into (10), then for a single low
frequency mode $i$ we obtain an estimate for its averaged energy
decay
$$
{{dE_i} \over {dt}} = - \left( {{2 \beta} \over {N}} \right) \omega_i
{{\beta E} \over {2\pi}} E_i E_h (t), \eqno(14)
$$
where from (3) $\omega_i = \pi i/N$.

Since $\delta k$ low frequency modes, assumed to have energy, couple
to $\delta h$ high frequency modes with $\delta k = \delta h$, the
cross-couplings imply each high frequency mode is coupled on average
to $\delta k/4$ low frequency modes. There are phases in the low
frequency mode beat oscillations and in the high frequency difference
oscillations that can affect the Arnold diffusion. This has only been
studied for exponentially slow diffusion [12]. The effect of these
phases when more than one driving term exists, for the case of
strong Arnold diffusion, $\Omega_B \,\, {\buildrel < \over \sim} \,\,
\delta \omega_h$, has not been studied. For lack of evidence we will
use the simplest assumption that the effect from each low frequency
mode is independent. Setting $\omega_i = \beta E/N$
$(i = \delta h/2 = \beta E / \pi )$ as an average value in (14), and dividing
by $E_i$, we obtain, an average, for each mode in the package
$$
{{dE_i} \over {E_i}} = - {{\beta} \over {\pi}} \,
\left( {{\beta E} \over {N}} \right)^2 E_h (t) dt, \eqno(15)
$$
with the assumption for scaling purposes that the number of couplings
is fixed. Integrating (15) in time, with $E_i (t)$ varying from
$E/\delta k$ at $t=0$ to the equipartition value $E/N$ at the final
time $T$ we get
$$
\ln \left( {{N} \over {\delta k}} \right) \simeq
\left( {{\beta} \over {\pi}} \right)
\left( {{\beta E} \over {N}} \right)^2
\int^T_0 E_h (t^{\prime}) dt^{\prime}\,, \eqno(16)
$$
Equation (15) only holds, initially, since $E$ decreases in time as
the diffusion proceeds. However, the change in $E$ is slow compared
to the initial build-up of the energy in the high frequency modes.
Furthermore, we expect that as the energy in the high frequency
modes increases, other pathways become available for the energy
distribution among the modes, to further justify holding the number
of couplings constant in the integration. The final step in the
approximation is to estimate the value of
$\int^T_0 E_h (t^{\prime})dt^{\prime}$ at $t=T$, a time of
``near-equipartition''. The quantity $E_h (t)$ appears in an integral,
so that its exact form is not required. For a diffusive process, in
which the amplitudes of the modes increase with $t^{1/2}$, we might
expect the mode energies to increase linearly with
$t$, $E_h (t) \simeq {t \over T} (E/N)$, such that the time dependence
does not depend on $N$. This is found to be approximately true,
numerically, over most of the evolution to near-equipartition. Other
forms of the time dependence of $E_h$ can also be taken with only
small numerical differences. Evaluating the integral with the
assumption of linear time dependence of $E_h (t)$ we obtain
$$
T \simeq {{2 \pi} \over {(\beta E/N)^3}} \ln
\left( {{\pi} \over {2 \beta E/N}} \right) . \eqno(17)
$$
We note that the logarithmic factor varies slowly. The
numerical coefficient is only a rough estimate. Equation (11) exhibits
a basic scaling of $T \propto (N/E)^3$. The scaling can be checked
numerically, by rescaling the time in Fig. 3. This is done in Fig. 4,
for five values of $E/N$, giving reasonable confirmation of the
rescaling of time with $(N/E)^3$. We also confirm this scaling by
plotting the time to reach $n_{eff}/N = 0.4$ against $E/N$, for all
of the data, comparing the result to the inverse cubic scaling
(dotted line), in Fig. 5. We can also compare the magnitude of $T$
in (17) with the numerics. Extrapolating the linear (with log time)
portion of the $E/N = 0.1$ curve in Fig. 3 to $n_{eff}/N = 1$ we
obtain, approximately, $T \sim 10^7$. Considering our many
approximations, this value is remarkably close to the value of
$T \simeq 3\cdot 10^7$ obtained from (17).

\centerline{\bf IV.\quad CONCLUSIONS AND DISCUSSION}

We have indicated, numerically, and justified, theoretically, that the
FPU-$\beta$ model has an appropriate thermodynamic limit. Provided
there is sufficient energy in a group of low frequency modes that
stochastic diffusion to high energy modes occurs on a
non-exponentially-slow time scale [1], then the dominant time scale
to equipartition is a power law $(N/E)^{\alpha}$.
The value  of $\alpha = 3$, estimated from a theoretical
scaling argument, was found to fit well to the numerical data.

The numerical results also clarify a result from a previous paper [2]
in which a $N^{1/2}$ scaling was numerically found, which would not
allow a finite-time thermodynamic limit. The resolution of the seeming
contradiction is that there is an initial transient which can extend
over much of the time to equipartition if $N$ is not very large and
the initial energy is placed in the first few modes (not proportional
to $N$). The existence of a thermodynamic limit also agrees with [4],
in which the energy was also placed in a mode packet
$\delta \gamma \propto N$. The power of $N/E$, in that study,
numerically fit better to $\alpha \simeq 1$. The use of a different
equipartition parameter, less sensitive than $n_{eff}/N$, could have
led to uncertainty in $\alpha$, but the issue has not been resolved.

We emphasize that the theory we have developed to explain the scaling,
does not predict the shape of $n_{eff}(t)$ which depends on
complicated dynamical processes. Furthermore, $n_{eff}$ is related
to the evolution of the energy in the individual modes in a very
complicated way. These dynamics lie beyond a simple mode-averaged
theory. We also emphasize that the theory depends on having
non-exponentially slow stochastic diffusion to high-frequency modes,
being driven by local mode-mixing stochasticity among low-frequency
modes [1]. For the approximations to be valid we require that $T \gg \tau$ where $\tau$ is the time scale for the assumed stochastic process.
Since $\tau \sim \Omega_B^{-1} \sim (N^2/\beta E)^2$,
the approximations hold if $\beta E/N \ll 1$.
\vfil\eject
\centerline{\bf ACKNOWLEDGMENTS}

We want to acknowledge the support from Fapesp, Brazil, (J.DeL.),
NSF Physics, USA, (A.J.L.), and INFN, Italy, (S.R.).
\vfil\eject
\centerline{\bf REFERENCES}

\item{[1]} J. De Luca, A. J. Lichtenberg, and M. A. Lieberman,
{\it Chaos}, {\bf 5,} 283 (1995).

\item{[2]} J. De Luca, A. J. Lichtenberg, and S. Ruffo,
{\it Phys. Rev. E}, {\bf 51,} 2877 (1995).

\item{[3]} J. De Luca, A. J. Lichtenberg, and S. Ruffo,
{\it Phys. Rev. E}, {\bf 54,} 2329 (1996).

\item{[4]} H. Kantz, R. Livi, and S. Ruffo, {\it J. Stat. Phys.},
{\bf 76}, 627 (1994).

\item{[5]} M. Pettini and M. Landolfi, Phys. Rev. A {\bf 41},
768 (1990).

\item{[6]} C. G. Goedde, A. J. Lichtenberg and M. A. Lieberman,
{\it Physica D}, {\bf 59}, 2000 (1992). 

\item{[7]} G. Tsaur and J. Wang, {\it Phys. Rev. E}, {\bf 54}, 4657
(1996).

\item{[8]} L. Casetti, M. Cerruti-Sola, M. Pettini, E. G. D. Cohen,
{\it Phys. Rev. E}, {\bf 55}, 6566 (1997).

\item{[9]} G. Parisi, {\it Europhys. Lett.}, {\bf 40},
357 (1997).

\item{[10]} S. Lepri, {\it Phys. Rev. E}, {\bf 58}, 7165 (1998).

\item{[11]} D.L. Shepelyansky, {\it Nonlinearity}, {\bf 10}, 1331
(1997)

\item{[12]} B. P. Wood, A. J. Lichtenberg, and M. A. Lieberman,
{\it Physica D}, {\bf 71}, 132 (1994).

\vfil\eject
\centerline{\bf FIGURES}

\item{\bf FIG. 1.} $n_{eff}/N$ vs time with $E/N=0.1$ and
$\delta \gamma = N/16$, for  $N$=256, 512, 1024, 2048 and 4096.
The error bars are the rms variation over (typically) 10 independent
trials, which are averaged to give final values.

\item{\bf FIG. 2.} $n_{eff}/N$ vs time with $E/N = 0.05$ and
$\delta \gamma = 4$ for $N$=256, 512, 1027, 2048.
Error bars as in Fig. 1.

\item{\bf FIG. 3.} $n_{eff}/N$ vs time for $N=2048$ and
$E/N$=0.03, 0.05, 0.1, 0.2, 0.4, and 0.8.

\item{\bf FIG. 4.} $n_{eff}/N$ vs $t(\beta E/N)^3$.

\item{\bf FIG. 5.} $t(n_{eff}/N = 0.4)$ vs $E/N$ compared to
proportionality $t$ vs $(E/N)^3$ of dashed line.

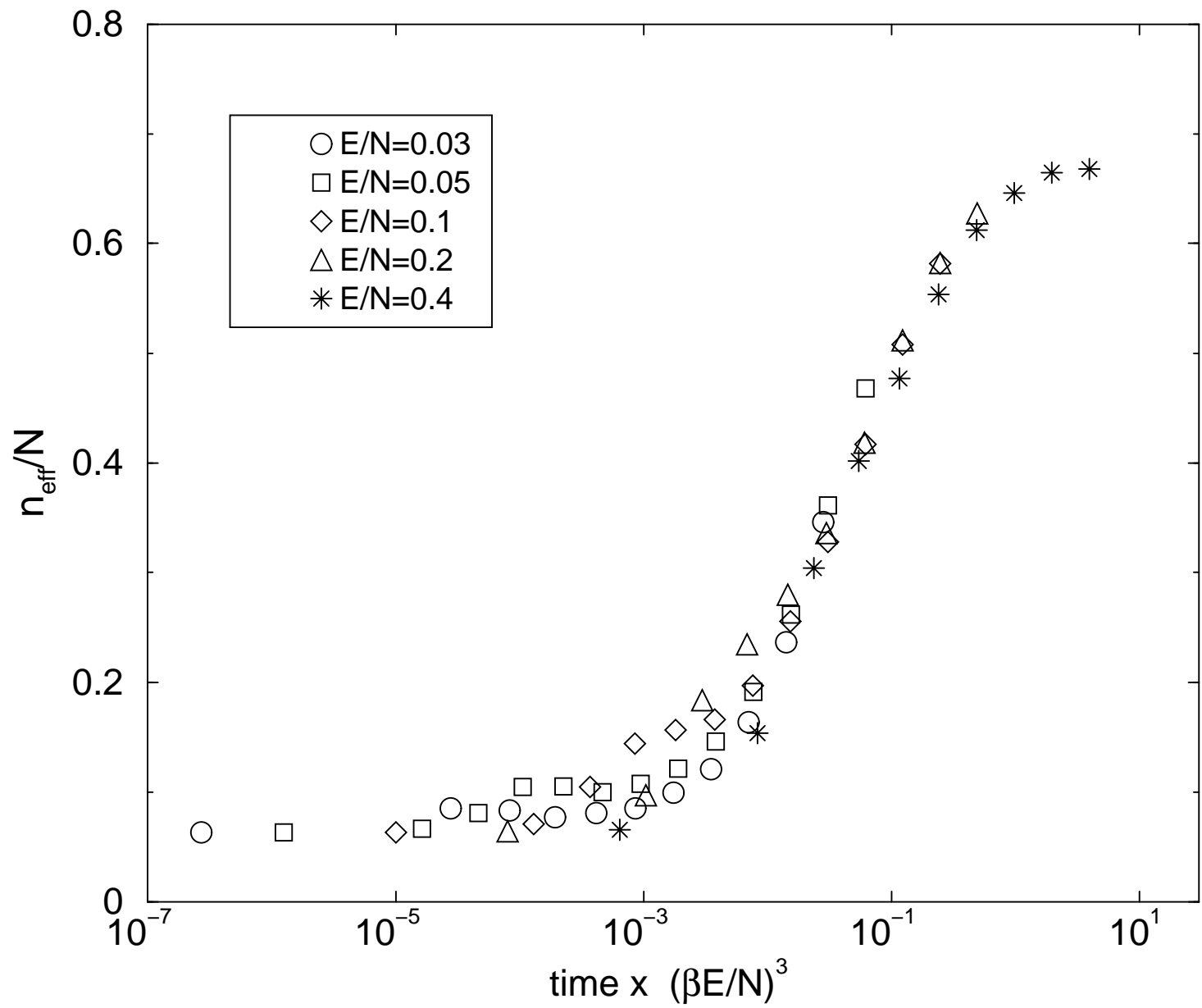
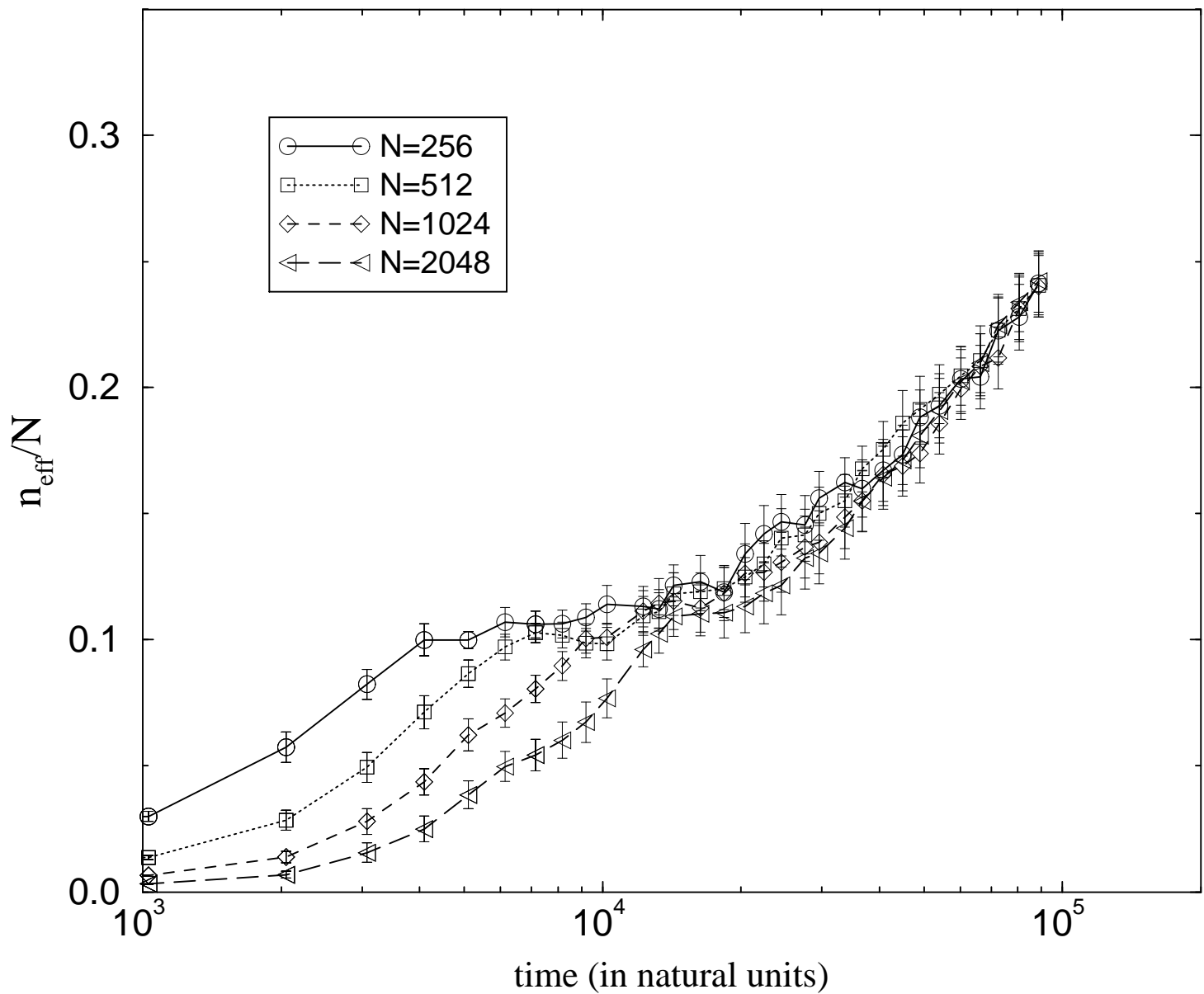
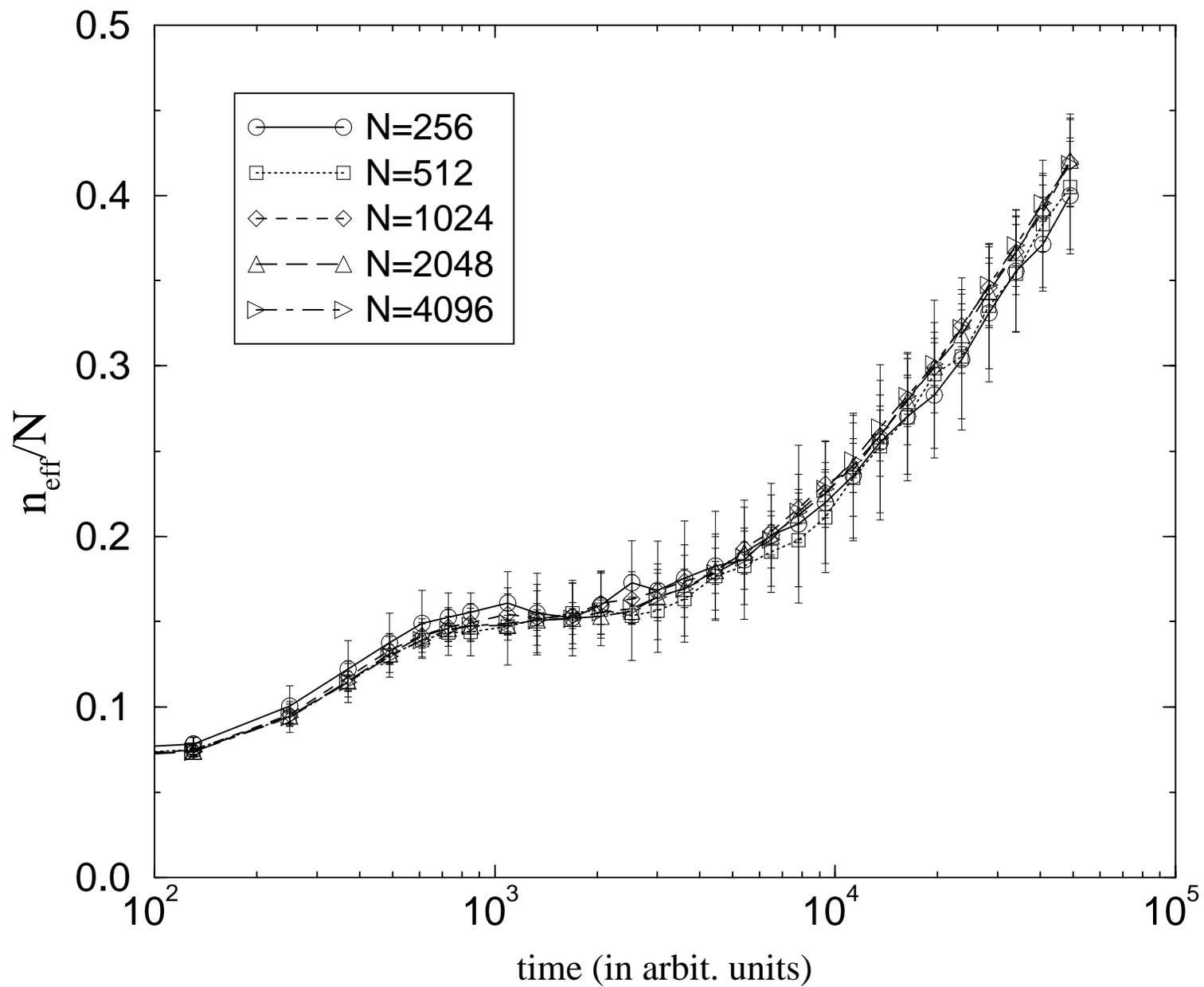
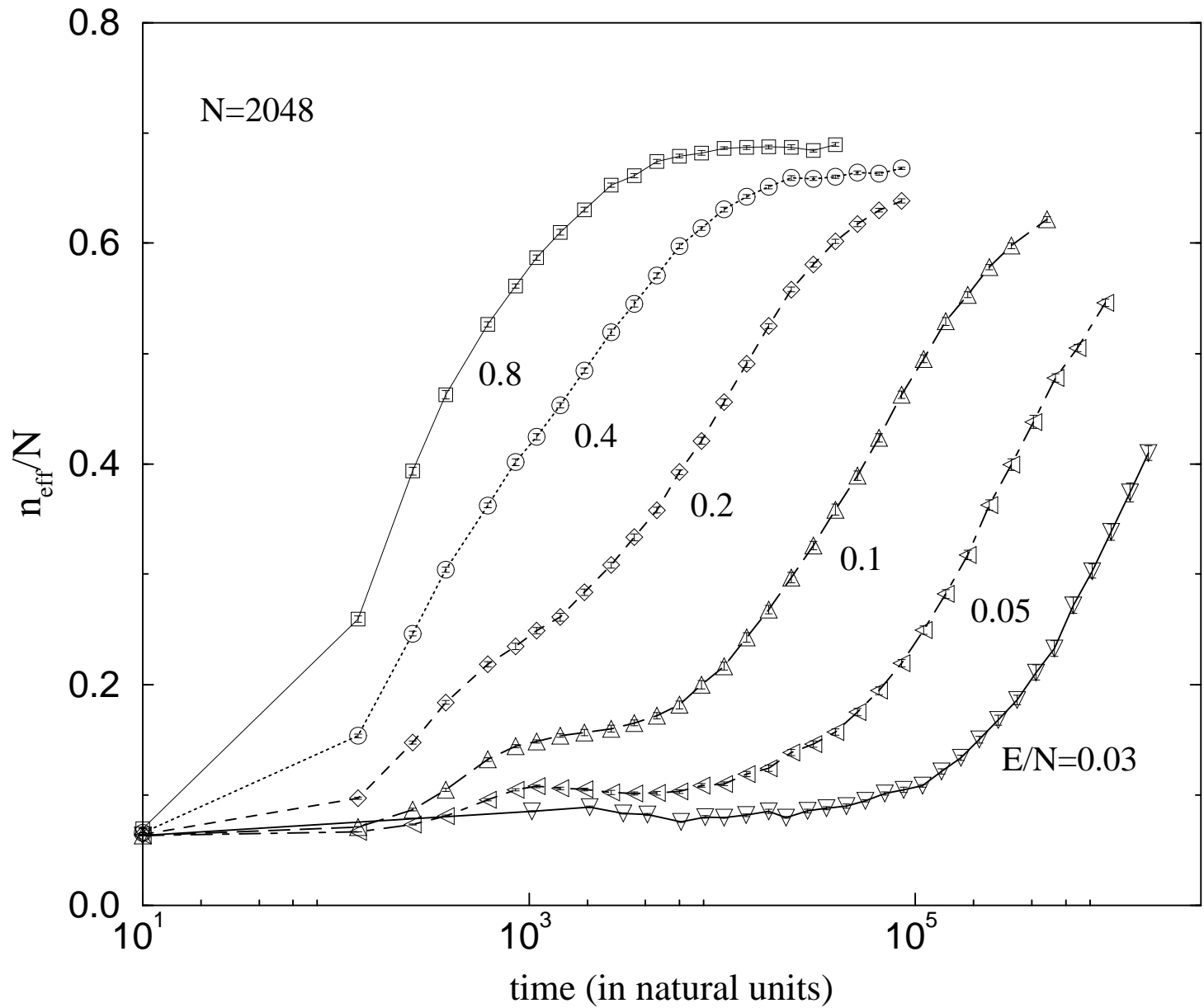
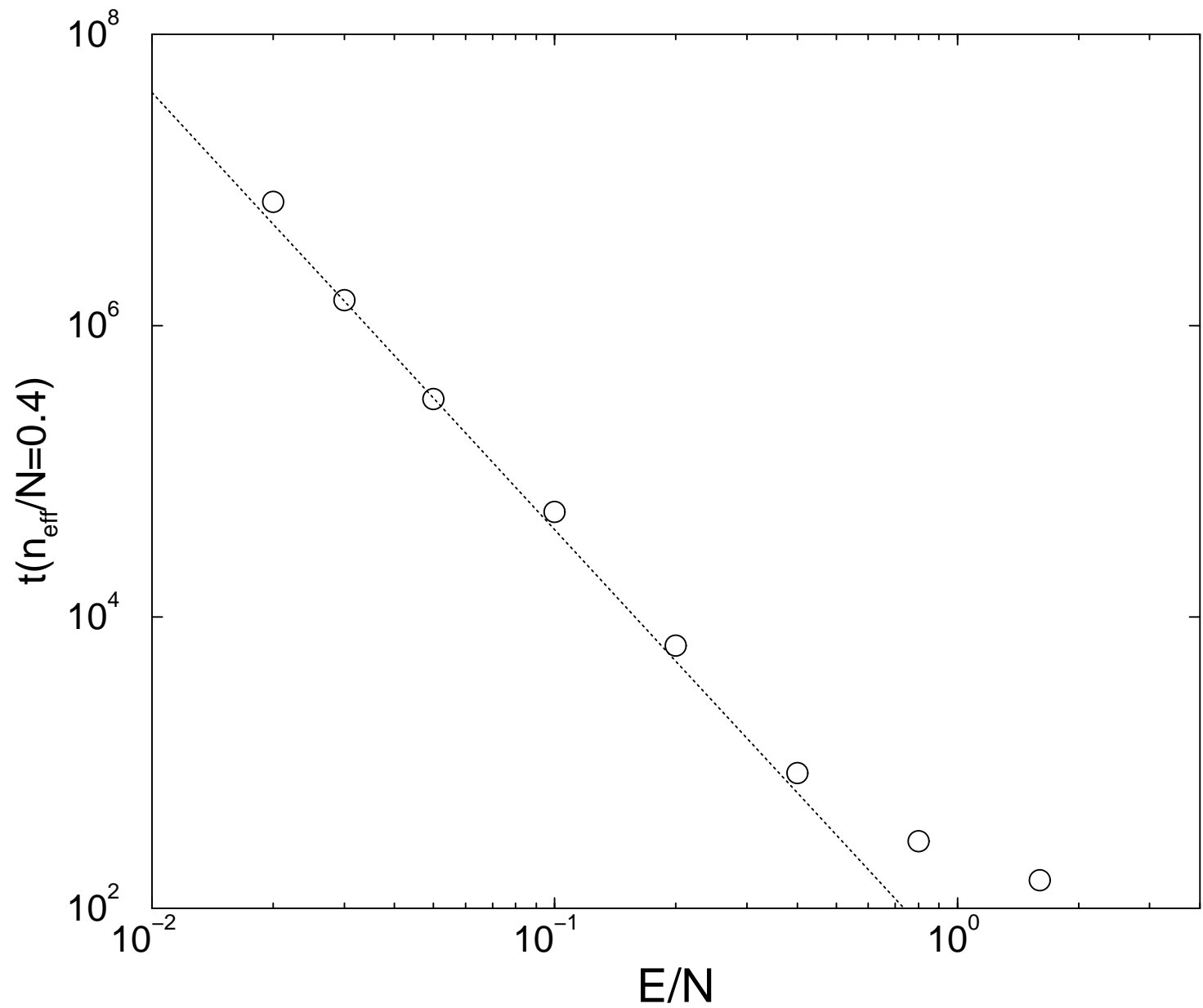
\end{document}